\documentclass[conference]{IEEEtran}
\usepackage{amsmath, amssymb, bm, cite, epsfig, psfrag}
\usepackage{graphicx}
\usepackage{soul} 
\usepackage[top    = 0.75in,
bottom = 1in,
left   = 0.63in,
right  = 0.63in]{geometry}

\usepackage{dblfloatfix}
\usepackage{array}
\usepackage{lipsum}

\usepackage[font=small]{caption}
\usepackage{epstopdf}	
\usepackage{longtable}
\usepackage{supertabular,booktabs}
\usepackage{bbm}
\usepackage{multirow}
\usepackage[usenames,dvipsnames]{xcolor}

\usepackage{subfigure}
\usepackage{etoolbox}
\usepackage{pbox}
\usepackage{fixltx2e}
\usepackage{tabu}	
\usepackage{filecontents}
\usepackage{enumerate}
\usepackage{textcomp}
\usepackage{colortbl}
\usepackage{fancyhdr}
\usepackage{bm}

\newtoggle{conference}

\togglefalse{conference} 
\graphicspath{{figures/}}

\newcolumntype{?}{!{\vrule width 2pt}}

\newcolumntype{P}[1]{>{\centering\hspace{0pt}}p{#1}}
\newcolumntype{M}[1]{>{\centering\hspace{0pt}}m{#1}}
\newcolumntype{L}{>{\centering\arraybackslash}m{3cm}}

\fancypagestyle{firststyle}{
	\fancyhf{}
	\fancyhead[L]{ \small Y. Xing, O. Kanhere, S. Ju, and T. S. Rappaport, ``Sub-Terahertz Wireless Coverage Analysis at 142 GHz in Urban Microcell,'' in  \textit{IEEE International Conference on Communications}, Seoul, South Korea, May 2022, pp. 1-6.}     

}
\IEEEoverridecommandlockouts

\begin{document}
\title{Sub-Terahertz Wireless Coverage Analysis at 142 GHz in Urban Microcell}
\author{\IEEEauthorblockN{ Yunchou Xing, Ojas Kanhere, Shihao Ju, and Theodore S. Rappaport}
	
	\IEEEauthorblockA{\small NYU WIRELESS, NYU Tandon School of Engineering, Brooklyn, NY, 11201,\\ \{ychou, ojask, shao, tsr\}@nyu.edu\\}
	
	\thanks{This research is supported by the NYU WIRELESS Industrial Affiliates Program and National Science Foundation (NSF) Research Grants: 1909206 and 2037845.}}

\maketitle
\thispagestyle{firststyle}

\begin{abstract}
Small-cell cellular base stations are going to be used for mmWave and sub-THz communication systems to provide multi-Gbps data rates and reliable coverage to mobile users. This paper analyzes the base station coverage of sub-THz communication systems and the system performance in terms of spectral efficiency through Monte Carlo simulations for both single-cell and multi-cell cases. The simulations are based on realistic channel models derived from outdoor field measurements at 142 GHz in urban microcell (UMi) environments conducted in downtown Brooklyn, New York. The single-cell base station can provide a downlink coverage area with a radius of 200 m and the 7-cell system can provide a downlink coverage area with a radius of 400 m at 142 GHz. Using a 1 GHz downlink bandwidth and 100 MHz uplink bandwidth, the 7-cell system can provide about 4.5 Gbps downlink average data rate and 410 Mbps uplink average data rate at 142 GHz.

\end{abstract}
    

\section{Introduction}~\label{sec:intro}

Radio frequencies above 100 GHz (e.g., sub-THz) are promising candidates for future communication systems to provide multi-Gbps average data rates, rapid streaming (e.g., the peak data rate on the order of Tbps) with extremely low latency (e.g., less than 1 ms), fiber-like fronthaul and backhaul in rural areas for fiber replacement and edge data centers replacements \cite{rappaport19access,Elayan20survey,viswanathan20A, ghosh195g}. Spectrum regulators (e.g., Japanese MIAC, European CEPT, and FCC in the US) have instituted their views and provisions on frequencies above 100 GHz during the past few years, and possible coexistence and spectrum sharing techniques above 100 GHz are presented in \cite{xing21a,marcus21a}.

The smaller wavelength (e.g., 1 mm) and wider available bandwidth (e.g., 1-40 GHz) at frequencies above 100 GHz will enable new techniques and applications like centimeter level precise position location \cite{Kanhere20a,Kanhere19a}, wireless cognition (e.g., robotic control, human surrogate) \cite{rappaport19access,viswanathan20A,zhang20a}, and joint communication and sensing \cite{ali20sensing,dokhanchi19a}. Learning the characteristics of the wireless propagation channel at frequencies above 100 GHz is the initial step for researchers to design future communications systems to realize the aforementioned applications. 

Many universities and research centers all over the world have conducted channel sounding measurements and research at frequencies above 100 GHz, including Aalto University \cite{nguyen2017comparing},  New York University (NYU) \cite{xing21icc}, University of Southern California (USC) \cite{abbasi20ICC}, Beijing Jiao Tong University (BJTU) \cite{zhang20a}, Shanghai Jiao Tong University (SJTU) \cite{han21icc}, Georgia Institute of Technology \cite{Kim17a}, Brown University \cite{ma18channel}, Technische University Braunschweig (TUBS) \cite{priebe11channel}, Koc University \cite{khalid16wideband}, University of Surrey \cite{demos20a}, and Pohang University of Science and Technology \cite{song20a}. These works have proven that sub-THz and THz frequencies can be used for future 6G wireless communications as the current mmWave 5G networks. However, what is the coverage for sub-THz communication systems and how is the system performance is still a question mark.

This paper analyzes the small-cell base station coverage and the system performance in terms of spectral efficiency (SE) at sub-THz frequencies in an Urban Microcell (UMi) environment, using realistic channel models derived from field measurements at 142 GHz conducted in downtown Brooklyn, New York \cite{xing21icc}. Section \ref{sec:SM} introduces the system model, channel models, and simulation settings. System performance results of single-cell and multi-cell cases for both uplink (UL) and downlink (DL) are evaluated in Section \ref{sec:results}. Finally, concluding remarks are drawn in Section \ref{sec:conclusion}.

\section{System Settings and Propagation Models}\label{sec:SM}
Small-cell architecture (shrinking the cell size from a few kilometers to a few hundreds of meters) is a promising means to increase area spectral efficiency and energy efficiency in future sub-THz and THz communications \cite{Rap17a,murdock2014consumption}. In this paper, small cell coverage is studied for both single-cell and multi-cell cases (7 cells in particular) in UMi scenarios at 142 GHz.

\begin{figure}    
	\centering
	\includegraphics[width=0.450\textwidth]{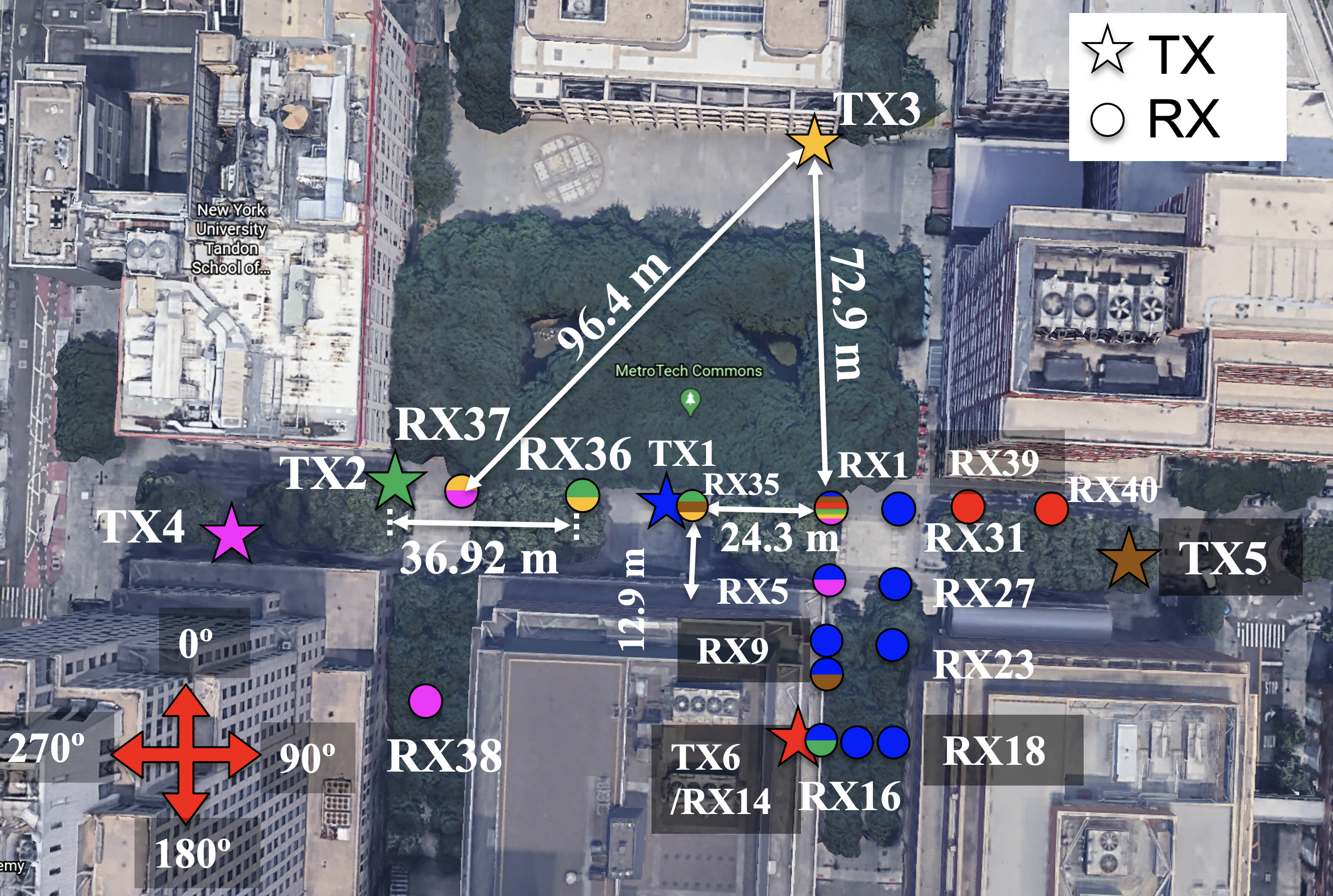}
	\caption{Terrestrial urban Microcell measurement campaign in NYU's downtown Brooklyn campus. Six TX locations are identified as stars with different colors and the corresponding RX locations are identified as the same color circles \cite{xing21icc}. }
	\label{fig:Mea1Loc}
\end{figure}

The propagation model and simulations in this paper are based on the outdoor wideband wireless propagation measurements at 142 GHz conducted in NYU downtown Brooklyn campus, which is a multipath-rich urban environment \cite{xing21icc}. The measurements were conducted at sea level in clear weather and standard air condition. Fig. \ref{fig:Mea1Loc} shows the measurement map with six TX locations and 17 RX locations (with some RX locations reused for more than one TX locations, such as RX1), resulting 16 LOS 16 LOS TX-RX location combinations and 12 NLOS TX-RX location combinations with TX-RX separation distances up to 117.4 m \cite{xing21icc}.   

Although atmospheric absorption and rain attenuation beyond the natural Friis free space loss at sub-THz and THz transmissions are greater than frequencies below 6 GHz (where air causes attenuation of only fractions of dB/km) \cite{rappaport19access,ITU-Rattenuation}, work in \cite{xing21a} shows that the atmospheric absorption has remarkably little impact on total path loss out to a few hundred meters and heavy rain will practically limit fixed THz links to several km. The atmospheric absorption at sea level in standard condition at 200-300 GHz is less than 10 dB/km, and even at 800-900 GHz the additional atmospheric absorption beyond the natural Friis free space loss is 100 dB/km at sea level, which is only 10 dB per 100 m over today’s 4G cellular, which will be compensated for by the antenna gains at higher frequencies \cite{xing21a}.   

In the simulation, the base stations (BS) are set at 4 m above the ground, working as small-cell lamppost BS (the same as measurements \cite{xing21icc}). The carrier frequency $f_c$ is at 142 GHz with 1 GHz null-to-null bandwidth, since 1 GHz to a few GHz bandwidths will be used for future sub-THz and THz communications \cite{xing21a,rappaport19access}. The user equipment (UEs) are set at 1.5 m above the ground with 15 dBi gain antennas and a 7 dB noise figure, as shown in Table \ref{tab:settings}.

\begin{table}[]
	\caption{System parameters for uplink (UL) and downlink (DL) propagation.}~\label{tab:settings}
	\begin{tabular}{|l|l|}
		\hline
		\textbf{Environment}           & Outdoor Urban Microcell (UMi) area \\ \hline
		\textbf{BS Height}             & 4.0 m above the ground             \\ \hline
		\textbf{Carrier Frequency}     & DL: 142 GHz, UL: 140 GHz                            \\ \hline
		\textbf{DL Channel Bandwidth}     & 1 GHz null-to-null                 \\ \hline
		\textbf{UL Channel Bandwidth}     & 100 MHz null-to-null                 \\ \hline
		\textbf{LOS Probability Model} & NYU (squared) Model (5GCM)         \\ \hline
		\textbf{Path Loss Model} & \begin{tabular}[c]{@{}l@{}}NYU CI Path Loss Model (n, $\sigma$):\\ Directional LOS: (2.1, 2.8 dB)\\ Directional $\text{NLOS}_{\text{Best}}$ (3.1, 8.3 dB)\end{tabular} \\ \hline
		\textbf{BS Transmit Power}     & 15 dBm (DL)                            \\ \hline
		\textbf{BS Antenna Gain}       & 40 dBi                             \\ \hline
		\textbf{BS Noise Figure}       & 5 dB                               \\ \hline
		\textbf{UE Transmit Power}     & 0 dBm (UL)                            \\ \hline
		\textbf{UE Antenna Gain}       & 15 dBi                             \\ \hline
		\textbf{UE Noise Figure}       & 7 dB                               \\ \hline
		\textbf{Single-cell UE Distribution}       & 250 UEs uniformly distributed     \\ \hline
		\textbf{7-cell UE Distribution}       & 1000 UEs uniformly distributed     \\ \hline
	\end{tabular}
\end{table}

The 1 m close-in (CI) free space reference distance path loss model \eqref{equ:CI} (with parameters derived from 142 GHz outdoor field measurements in a UMi environment in downtown Brooklyn \cite {xing21icc, Xing21c}) is used for the link budget calculation:
\begin{equation}
	\label{equ:CI}
	\small
	\begin{split}
		PL^{CI}(f_c,d_{\text{3D}})\;\text{[dB]} &= \text{FSPL}(f_c, 1 m) +10n\log_{10}\left( \dfrac{d_{3D}}{d_{0}} \right)+ \chi_{\sigma},\\
		\text{FSPL}(f_c,1 m) &= 32.4 + 20\log_{10}(\dfrac{f_c}{1\;\text{GHz}}),
	\end{split}
\end{equation}
where FSPL$(f_c, 1 \;\text{m})$ is the large-scale free space path loss at carrier frequency $f_c = 142$ GHz at 1 m, $n$ is the path loss exponent (PLE), and $\chi_{\sigma}$ is the large-scale fading in dB (a zero mean Gaussian random variable with a standard deviation $\sigma$ in dB). The directional PLE and shadow fading $(n, \sigma)$ for LOS and $\text{NLOS}_{\text{Best}}$ directions (when the BS and UE antennas are pointing in the direction where the maximum power is received) are (2.1, 2.8 dB) and (3.1, 8.3 dB) at 142 GHz, respectively.

\begin{figure}    
	\centering
	\includegraphics[width=0.45\textwidth]{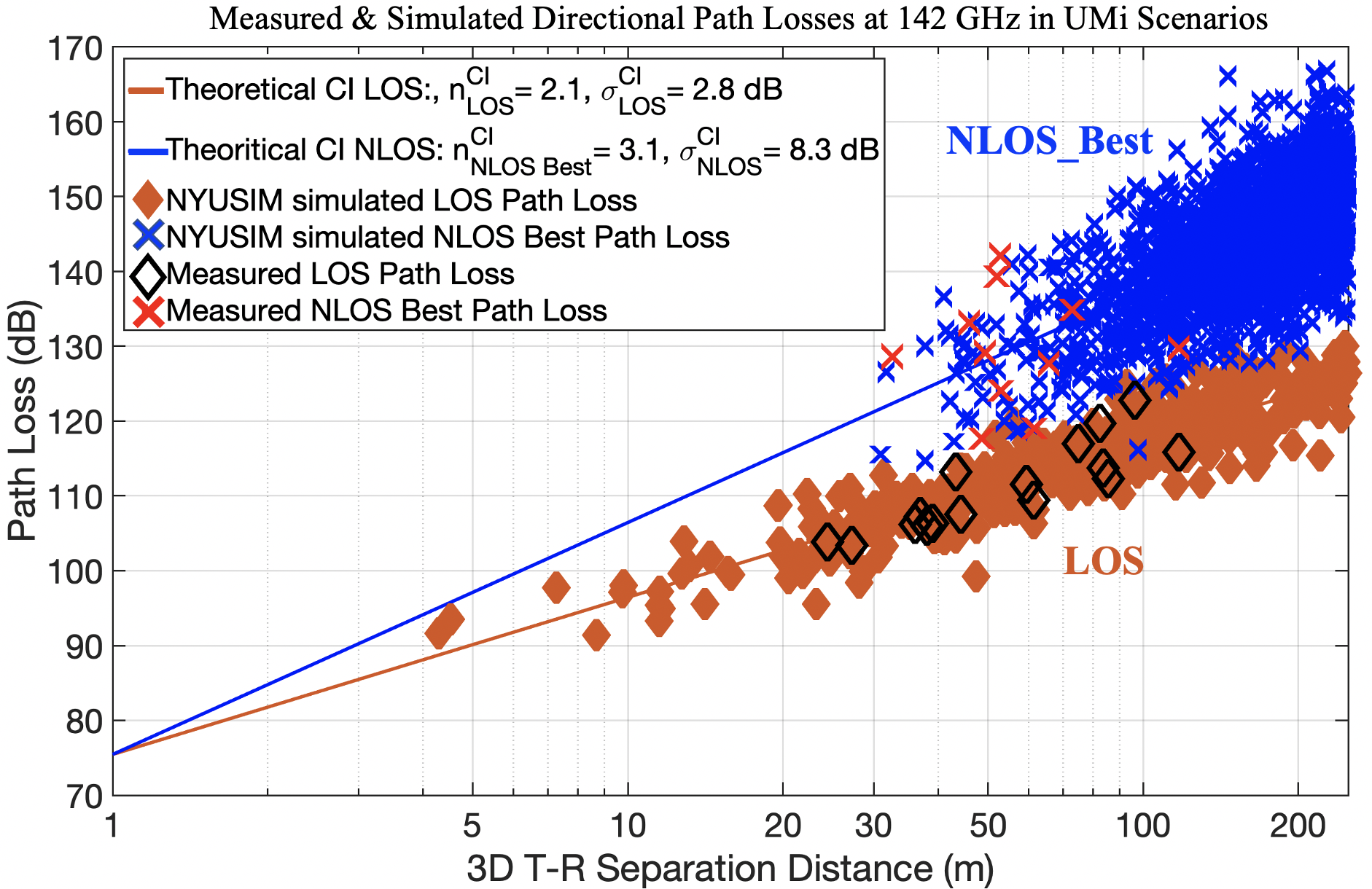}
	\caption{Measured and simulated directional path loss (with antenna gains removed) at 142 GHz in UMi environment. The LOS and NLOS Best path loss data (black diamonds and red crosses) are from outdoor field measurements at 142 GHz in downtown Brooklyn \cite{xing21icc,Xing21b}, with a standard deviation of 2.8 dB and 8.3 dB to the theoretical CI model \eqref{equ:CI} for LOS and NLOS cases, respectively. The simulated path loss data are generated by NYUSIM \cite{NYUSIM}, with a standard deviation of 2.7 dB and 8.5 dB to the theoretical CI model \eqref{equ:CI} for LOS and NLOS cases, respectively.}
	\label{fig:PL142GHz}
\end{figure}

The theoretical CI path loss models and measured path loss data of outdoor field measurements at 142 GHz \cite{xing21icc,Xing21b} in downtown Brooklyn, New York, are presented in Fig.\ref{fig:PL142GHz}. Additional simulated path loss data are generated by NYUSIM \cite{NYUSIM} which is an open source channel simulator based on field measurements from mmWave to sub-THz frequencies.

For radio propagation at frequencies above 100 GHz, there is still a myth that the wireless channels at higher frequencies would experience more path loss as only omni-directional antennas are considered at both the link ends \cite{rappaport19access}. If the effective aperture is kept constant over frequencies at both the TX and RX, the path loss decrease quadratically as frequency increases, although the current antenna techniques may not be able to keep the effective aperture constant at sub-THz frequencies compared to mmWave or lower frequencies. 

Work in \cite{Xing21b,Xing21c} shows that path loss exponents (PLEs) are remarkably similar over frequencies from 28 GHz to 142 GHz, when referenced to the first meter free space path loss. This means after the first meter propagation, wireless channels at sub-THz frequencies do not provide more loss to signals than the channels at mmWave frequencies. Measurements at 142 GHz have shown that metal surfaces, metal lampposts, glass walls, concrete walls, and marble pillars are good reflectors at sub-THz frequencies. Additionally, work in \cite{Ju21Globecom,Ju21a} shows that 1-6 spatial clusters observed at 142 GHz with a 20 dB threshold form the peak for both LOS and NLOS scenarios in an UMi environment, and the number of spatial clusters follows Poisson distributions (statistics are sensitive to noise threshold).

\begin{figure}    
	\centering
	\includegraphics[width=0.45\textwidth]{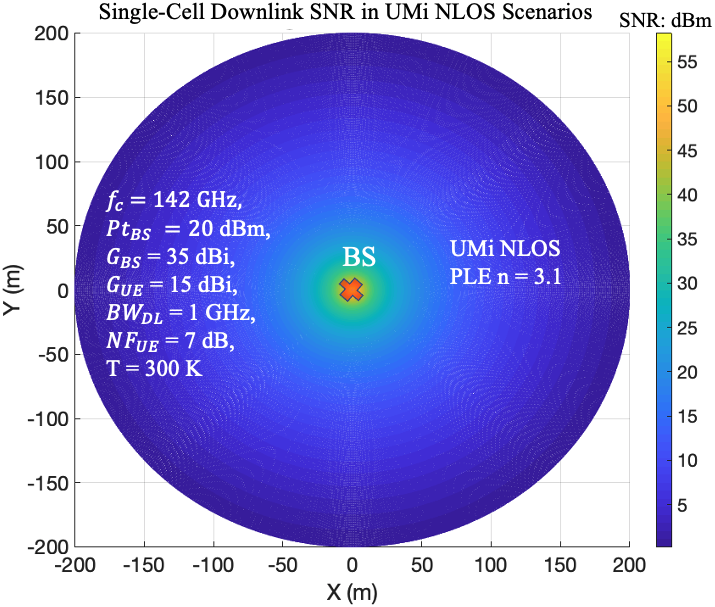}
	\caption{Single-Cell downlink coverage with SNR larger than 0 dB in UMi NLOS scenarios, assuming both BS and UE point to the $\text{NLOS}_{\text{Best}}$ direction.}
	\label{fig:SingleCellCoverage}
\end{figure}

\begin{figure}    
	\centering
	\includegraphics[width=0.45\textwidth]{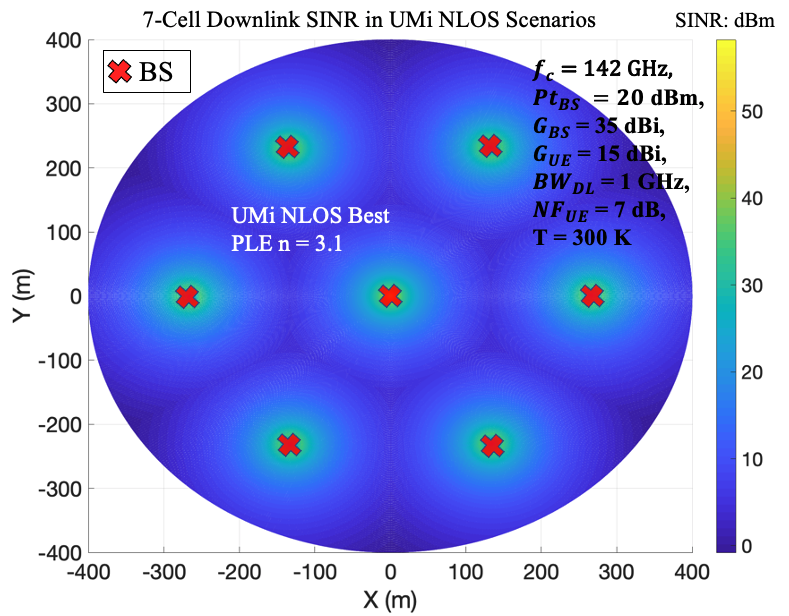}
	\caption{7-Cell downlink coverage with SINR larger than 0 dB in UMi NLOS scenarios, assuming the UE points to the $\text{NLOS}_{\text{Best}}$ direction of the closest BS, and the signal powers from other BS are considered as interference.}
	\label{fig:7CellCoverage}
\end{figure}

In our simulation, the BS EIRP (effective isotropic radiated power) is set the same as the current BS working at low mmWave frequencies (e.g., 28 GHz with an EIRP of 55 dBm), which is smaller than the maximum EIRP regulated by FCC (e.g., maximum average EIRP of 60 dBm for BS and 40 dBm for mobile). The BS is set at the origin in the single-cell case with a radius of 200 m coverage for a signal to noise ratio (SNR) larger than 0 dB in UMi NLOS scenarios. Fig. \ref{fig:SingleCellCoverage} shows the single-cell downlink coverage calculated using \eqref{equ:CI} with PLE $n = 3.1$ for $\text{NLOS}_{\text{Best}}$ scenarios (average path loss without fading). A single base station with an EIRP  of 55 dBm at 142 GHz can cover an area with a radius of 200 m even in NLOS scenarios.

For the multi-cell case (7-cell in particular), a BS is set at the origin with six other BS equally spaced, as shown in Fig. \ref{fig:7CellCoverage}. Considering NLOS scenarios, the UEs point to the $\text{NLOS}_{\text{Best}}$ direction where the maximum power is received, and the signal powers from other BS are considered as interference. It shows that 7-cell architecture can provide a coverage area (with SINR larger than 0 dB) with a radius of 400 m in NLOS scenarios. The LOS coverage areas of both single-cell and 7-cell architectures are much larger than in NLOS environments. Thus, to study the coverage and system performance at 142 GHz, 1000 UEs are uniformly distributed in areas with a radius of 200 m and 400 m for single-cell and 7-cell cases, respectively.

The NYU (squared) LOS probability model \eqref{equ:LOS} (a statistical model derived from a real-world database in downtown New York City \cite{A5GCM15,Rap17a}) is used to predict whether the UE is within a clear LOS of the BS or in an NLOS region due to obstructions:
\begin{equation}
	\label{equ:LOS}
	\begin{split}
	P_{LOS} (d_{2D}) = & (\min(d_1/d_{2D}, 1)(1-\exp(-d_{2D}/d_2)) \\
	& +\exp(-d_{2D}/d_2))^2,
	\end{split}
\end{equation}
where $P_{LOS} (d_{2D})$ is the likelihood that a UE is in a clear LOS of the BS, $d_{2D}$ is the 2D Euclidean separation distance between the BS and UE in meters, $d_1 = 22$ m, and $d_2 = 100$ m.  

\begin{figure}    
	\centering
	\includegraphics[width=0.45\textwidth]{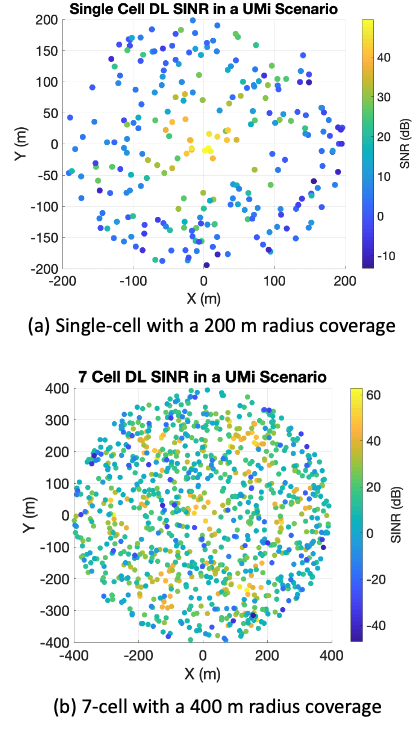}
	\caption{UE distributions for single-cell and 7-cell scenarios.}
	\label{fig:UElocations}
\end{figure}

Fig. \ref{fig:UElocations} shows the distribution of UE for both single-cell and 7-cell scenarios, and different colors of UEs representing the SNR level with dark blue corresponding to outage (with SNR less than -10 dB). The LOS status of each UE is determined by \eqref{equ:LOS} and about 20\% of the UE is in LOS of the BS in the single-cell case, as shown in Fig. \ref{fig:UElocations}(a). It is worth noting that, in the 7-cell case, a UE is determined the LOS status of each BS based on the BS-UE separation distance using \eqref{equ:LOS}, and about 37\% of UEs are in LOS of at least one BS, as shown in Fig. \ref{fig:UElocations}(b). 

In the single-cell case, the UEs in LOS of the BS point to the LOS boresight directions to the BS and the UEs in NLOS region point to the $\text{NLOS}_{\text{Best}}$ directions. The downlink (DL) received power is calculated by: 
\begin{equation}
	\label{equ:SNR}
	\begin{split}
		Pr_{dl} = Pt_{bs} + G_{bs} + G_{ue} - 	PL^{CI}(f_c,d_{\text{3D}}),\\
	\end{split}
\end{equation} 
where the PLE and shadow fading of (2.1, 2.8 dB) and (3.1, 8.3 dB) are used for directional path loss models in LOS and $\text{NLOS}_{\text{Best}}$ scenarios. As shown in Fig. \ref{fig:UElocations}(a), there are ~10\% of the UEs are out of coverage (with an SNR lower than 0 dB).

In the multi-cell case, the UEs point to the directions where the maximum powers are received (i.e., the LOS boresight directions or the strongest $\text{NLOS}_{\text{Best}}$ directions to a certain BS). The signal powers from the rest six BS are considered as interference. As shown in Fig. \ref{fig:UElocations}(b), there are ~15\% of the UEs are out of coverage (with an SINR lower than 0 dB).

It is worth noting that the propagation model used in this paper are based on field measurements conducted in an UMi environment at sea level in a clear weather and standard air condition, and additional effects (e.g., rain/snow/fog, different altitudes and humidity) can be added to the models for different scenarios.

\section{Simulation Results of Single Cell and Multicell Cases}\label{sec:results}

The spectral efficiency (SE) of data transmission is an important performance of communication systems, which usually varies across the service and depends on the UE locations as well as channel conditions \cite{xing21vtc}. The user SE is defined as the amount of data transferred to (DL) or from (UL) the user normalized by the time-frequency resources used by the system for data transfer. Suppose $M_{u}$ packets have been transmitted for user $u$, the SE $\eta_u$ is calculated as:
\begin{equation}
	\eta_u = \left( \sum_{i=1}^{M_u}N_{u,i} \right)/ \left(  \sum_{i=1}^{M_u}T_{u,i}B_{u,i}\right),  
\end{equation}
where $N_{u,i}, T_{u,i}, B_{u,i}$ are respectively the number of information bits, transmission time, and allocated bandwidth for the $i$-th packet of user $u$.

\begin{figure}    
	\centering
	\includegraphics[width=0.45\textwidth]{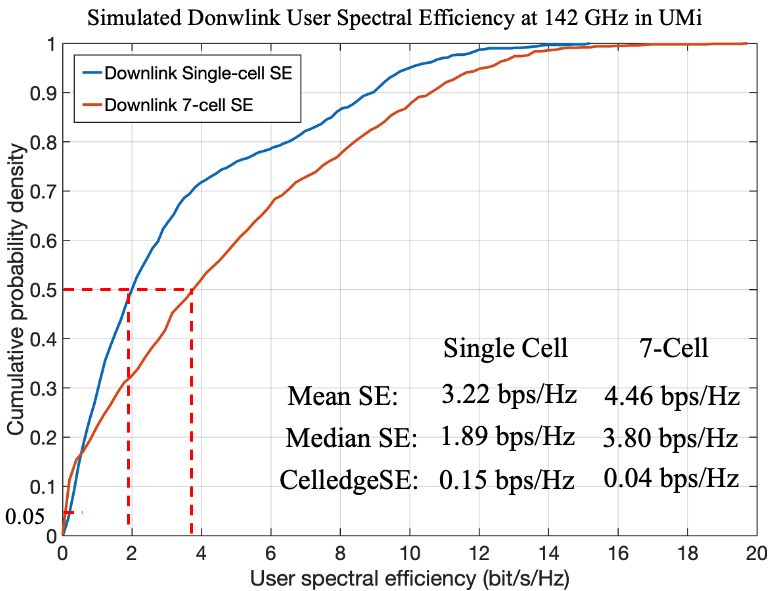}
	\caption{Simulated downlink User Spectral Efficiency at 142 GHz in UMi environment with a downlink channel bandwidth of 1 GHz.}
	\label{fig:SEDL}
\end{figure}

Fig. \ref{fig:SEDL} presents the downlink spectral efficiency (SE) distribution at 142 GHz for both the single-cell case (with a radius of 200 m) and 7-cell case (with a radius of 400 m) with the BS and UEs using directional antennas, as shown in Table \ref{tab:settings}. In the single-cell case (blue link in Fig. \ref{fig:SEDL}), the average user SE is 3.22 bps/Hz, the median user SE is 1.89 bps/Hz, and the cell-edge user (5\% of the users either at the edge of the cells or in deep fading area) SE is 0.15 bps/Hz. With a 1 GHz bandwidth, the average downlink data rate is 3.22 Gbps and the cell-edge users experience an average downlink date of 150 Mbps. The 7-cell architecture can increase the total coverage to a radius of 400 m (by adding six more BS and shrinking the cell size of each BS), with the average user SE of 4.46 bps/Hz, the median user SE of 3.80 bps/Hz, and the cell-edge user SE of 0.04 bps/Hz. With a 1 GHz bandwidth, the average downlink data rate for the 7-cell case is 4.46 Gbps and the cell-edge users experience an average downlink date of 40 Mbps.

It shows that using the 7-cell architecture can greatly improve the average (+40\%) and median user SE (+100\%) compared to the single-cell architecture, however, the cell-edge users experience worse service (-75\%) due to the interference.     

\begin{figure}    
	\centering
	\includegraphics[width=0.45\textwidth]{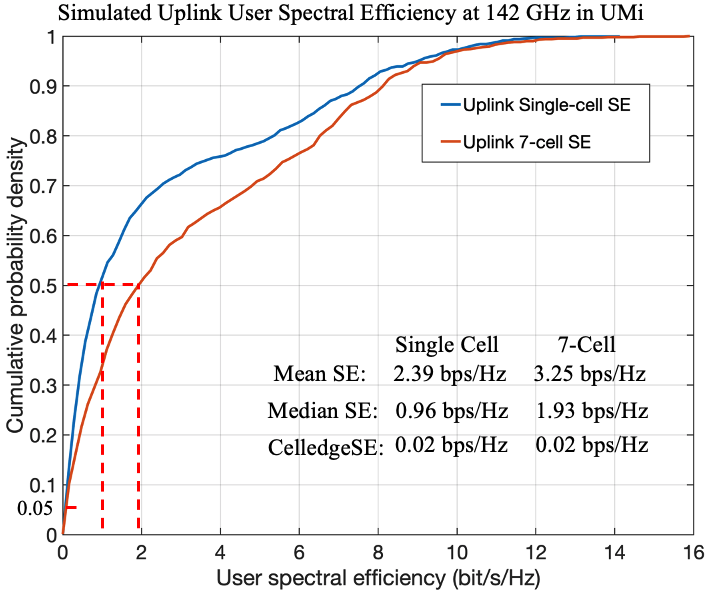}
	\caption{Simulated uplink user spectral efficiency at 140 GHz in UMi environment with an uplink channel bandwidth of 100 MHz.}
	\label{fig:SEUL}
\end{figure}

Fig. \ref{fig:SEUL} shows the uplink user spectral efficiency distribution at 142 GHz for both the single-cell case (with a radius of 200 m) and the 7-cell case (with a radius of 400 m) with the BS and UEs using directional antennas. In the single-cell case, the uplink SE decrease significantly compared to the downlink SEs, with the mean SE of 2.39 bps/Hz (-26\%), the median SE of 1.93 bps/Hz (-50\%), and the cell-edge SE of 0.02 bps/Hz (-87\%). The uplink coverage is limited by the low EIRP of the UE devices (15 dBm). With a 100 MHz bandwidth, the average data rate in the sing-cell case is 239 Mbps and the cell-edge data rate is 20 Mbps.

In the 7-cell case, the uplink SEs also decrease (compared to the downlink SEs) due to the limited UE EIRP, with the mean SE of 3.25 bps/Hz (-27\%), the median SE of 1.93 bps/Hz (-49\%), and the cell-edge SE of 0.02 bps/Hz (-50\%). In general, using the 7-cell architecture can greatly improve the average and median SE compared to using the single-cell architecture, but have a worse performance for the cell-edge users.

\section{Conclusion}\label{sec:conclusion}
In this paper, we have analyzed both the single-cell and multi-cell base station coverage of sub-THz systems at 142 GHz in outdoor UMi scenarios based on realistic channel models derived from field measurements. The single-cell base station can provide coverage of a 200 m radius, and the 7-cell base stations can provide total coverage of a 400 m radius. Uplink and downlink system performances in terms of spectral efficiency are studied for both single-cell and multi-cell cases. The 7-cell architecture can greatly improve the coverage and system performance in terms of the average and median SE compared to the single-cell case but has a worse performance to the cell-edge users. In the 7-cell case, with a 1 GHz bandwidth, DL sector throughput is 4.5 Gbps and DL cell-edge throughput is 40 Mbps. The uplink coverage is limited by UE transmit power and UE antenna gain. Dynamic blockages (e.g., human and vehicle blockages) and fast beam switching (when there is an obstruction block the link) algorithms will be studied for future work.

\bibliographystyle{IEEEtran}
\bibliography{Indoor140GHznew}

\end{document}